\documentstyle[amssymb,prl,aps,twocolumn,epsf]{revtex}

\begin{document}
\title{Out-of-equilibrium Kondo effect in Double Quantum Dots}
\author{Ram\'{o}n Aguado and David C. Langreth}
\address{
Center for Materials Theory, Department of Physics and Astronomy, Rutgers University, Piscataway, NJ
08854-8019, USA.}
\maketitle

\begin{abstract}
The out-of-equilibrium transport properties of a double quantum
dot system in the Kondo regime are studied theoretically by means
of a two-impurity Anderson hamiltonian with interimpurity hopping.
The hamiltonian is solved by means of a non-equilibrium generalization of the slave-boson mean-field theory.
It is demonstrated that measurements
of the differential conductance $dI/dV$,
for the appropriate values of voltages and tunneling couplings, can give a direct observation
of the coherent superposition between the many-body Kondo states of
each dot. For large voltages and arbitrarily large interdot tunneling, 
there is a critical voltage above which
the physical behaviour of the system 
again 
resembles
that of two decoupled quantum dots. 
\end{abstract}
\pacs{PACS numbers:XXX}
\tighten

\thispagestyle{empty}
\narrowtext
Recent experiments \cite{Gold,Leo,Stutt} have shown that new physics 
emerge when the
transport properties
of quantum dots (QD's) at temperatures (T) below the 
Kondo temperature ($T_K$) are studied.\cite{Kon} 
QD's offer the intriguing possibility of a continuous tuning of the relevant parameters
governing the Kondo effect \cite{Hew} as well as the possibility of studying Kondo physics
when the system is driven out of equilibrium in different ways.\cite{noneq}
These experimental breakthroughs have opened up a new way for the study of
strongly 
correlated electrons in artificial systems.
The Kondo anomaly appearing in the density of states (DOS) 
of the QD reflects the formation of a quantum-coherent many-body state. 
Motivated by the recent experimental advances in the study of
double quantum dots (DQD) \cite{Tosh} 
it is thus interesting to study what happens
when two QD's in the Kondo regime are coupled.
Previous theoretical studies of this problem at equilibrium 
have focused on the competition between Kondo effect and antiferromagnetic coupling generated
via superexchange \cite{Geor,Num} or via capacitive coupling between dots.\cite{Andr}\\
In this Letter we focus on the study of a DQD in the Kondo regime
driven out of equilibrium
by means of a
DC voltage bias.
There have hitherto been only few attemps to study this problem \cite{Ivan}
but a clear picture of the problem is yet missing.
Following the recent work of Aono {\it et al} \cite{Aon} 
and Georges and Meir \cite{Geor} 
we employ the slave boson (SB) technique \cite{slave} in a mean field approximation (MFA)
and generalize it to a non-equilibrium situation.
This MFA allows us to 
include nonperturbatively the interdot tunneling term (i.e coherence between dots).
The different physical regimes that appear as the ratio $\tau_c=t_C/\Gamma$ changes 
($t_C$ is the interdot tunneling coupling and $\Gamma$ is the
single particle broadening coming from the coupling to leads\cite{wide})
can be explored  by measuring the non-linear transport
properties of the system.
Our results can be summarized in Figs. 1 and 2:
the differential conductance dI/dV of the DQD 
directly measures the transition (as $\tau_c$ increases)
from two isolated Kondo impurities
to a coherent superposition of the many-body Kondo states of
each dot, which form bonding and anti-bonding combinations. 
This coherent state which occurs for $\tau_c>1$
is reflected as a splitting of the zero-bias anomaly in the dI/dV curves. 
This splitting depends non-trivially on the voltage and on the many-body 
parameters of the problem.
For large voltages, we find that
there is a critical voltage above which
the coherent configuration is unstable and
the physical behaviour of the system
again resembles that of two decoupled QD's, i.e two Kondo singularities pinned at each
chemical potential, even for $\tau_c>1$. This instability
is reflected as a drastic drop of the current leading
to singular regions of negative differential conductance (NDC).\\
{\it Model}: 
In typical experiments, $U_{\rm intradot},\Delta\epsilon>>T$ 
($U_{\rm intradot}$ is the
strong on-site Coulomb interaction on each dot, $\Delta\epsilon$ is the average level separation),
which allows one to consider a single state in each QD.\cite{multilevel}
We can model the DQD with a (N=2) fold degenerate two-impurity 
Anderson hamiltonian
with an extra term accounting for interdot tunneling.
Each impurity is coupled to a diferent Fermi sea of chemical 
potential $\mu_L$ and $\mu_R$, respectively.
In the limit $U_{\rm intradot}\rightarrow\infty$ (on each QD) and 
$U_{\rm interdot}\rightarrow 0$ \cite{inter}, 
the hamiltonian may be written
in terms of auxiliary SB 
operators \cite{slave} plus constraints:
\begin{eqnarray}
H&=&\sum _{ k_{\alpha\in \{ L,R \}}, \sigma} \epsilon _{k_\alpha} c^\dagger_{k_\alpha,\sigma}
c_{k_\alpha,\sigma}
+ \sum _{\alpha\in \{ L,R \},\sigma} \epsilon _{\alpha\sigma} f^\dagger_{\alpha\sigma}
f_{\alpha\sigma}\nonumber\\
&+&\frac{t_C}{N}\sum _\sigma (f^\dagger_{L\sigma}b_L b^\dagger_R f_{R\sigma}
+f^\dagger_{R\sigma}b_R b^\dagger_L f_{L\sigma})\nonumber\\
&+&\frac{1}{\sqrt N}\sum _{k_{\alpha\in \{ L,R \}},\sigma }V_{\alpha}(c^\dagger_{k_\alpha, \sigma} b^\dagger_\alpha 
f_{\alpha\sigma}
+ f^\dagger_{\alpha\sigma} b_\alpha c_{k_\alpha, \sigma})\nonumber\\
&+&\sum_{\alpha\in \{ L,R \}} \lambda_\alpha(\sum_\sigma f^\dagger_{\alpha\sigma}f_{\alpha\sigma}
+b^\dagger_\alpha b_\alpha-1).
\end{eqnarray}
$c^\dagger_{k_\alpha,\sigma}(c_{k_\alpha,\sigma})$ are the creation (annihilation) operators
for electrons in the lead $\alpha$. To simplify the notation we consider 
henceforth that $V_L=V_R=V_0$ and $\epsilon _{L\sigma}=\epsilon _{R\sigma}=\epsilon _{0}$ 
(i.e, $T_K$ is the same for both dots at equilibrium. 
The generalization to different $T_K$'s is straightforward). 
The even-odd symmetry 
is broken by the interdot coupling $t_C$.
In the SB representation, the annihilation operator for electrons in the QD's, 
$c_{\alpha\sigma}$ is decomposed into the SB operator $b^\dagger_\alpha$ 
which creates an empty state and a pseudo fermion operator $f_{\alpha\sigma}$ which annihilates
the singly occupied state with spin $\sigma$ in the dot $\alpha$: 
$c_{\alpha\sigma}\rightarrow b^\dagger_\alpha f_{\alpha\sigma}$ 
($c^\dagger_{\alpha\sigma}\rightarrow f^\dagger_{\alpha\sigma}b_\alpha$). 
In the last term of (1), the charge operator 
$\hat{Q}_\alpha=\sum_\sigma f^\dagger_{\alpha\sigma}f_{\alpha\sigma}+b^\dagger_\alpha b_\alpha$ 
has been introduced. 
This term represents the constraint $\hat{Q}_\alpha=1$ in each dot with Lagrange multiplier
$\lambda_\alpha$. This constraint prevents double occupancy 
in the limit $U\rightarrow\infty$. \\ 
{\it Solution}: In the lowest order, we assume that the SB operator 
is a constant c-number
$b_\alpha(t)/\sqrt N=\langle b_\alpha(t)\rangle/\sqrt N=\tilde{b}_\alpha$ neglecting the 
fluctuations around the average $\langle b_\alpha(t)\rangle$ of
the SB. At T=0, this MFA is correct for describing
spin fluctuations (Kondo regime). Mixed-Valence
behavior (characterized by strong charge fluctuations) cannot be described
by the MFA. This restricts our non-equilibrium calculation to low
voltages $V<<\epsilon_{0}$.
Charge fluctuations can be included as thermal or quantum fluctuations 
($1/N$ corrections). \cite{slave,Gauss}
Defining $\tilde{V}_\alpha=V_0\tilde{b}_\alpha$ and $\tilde{t}_C=t_C\tilde{b}_L\tilde{b}_R$ 
we obtain from the constraints and the equation of motion of the SB operators 
the selfconsistent set of four equations with four unknowns 
($\tilde{b}_L,\tilde{b}_R,\lambda_L,\lambda_R$):
\begin{eqnarray}
&& \tilde{b}_{L(R)}^2+\frac{1}{N}\sum_\sigma 
\langle f^\dagger_{L(R)\sigma}f_{L(R)\sigma}\rangle=\frac{1}{N}\nonumber\\
&&\frac{\tilde{V}_{L(R)}}{N}\sum _{k_{L(R)},\sigma }\langle c^\dagger_{k_{L(R)}, \sigma} 
f_{L(R)\sigma}\rangle\nonumber\\
&&+\frac{\tilde{t}_C}{N}\sum_\sigma 
\langle f^\dagger_{R(L)\sigma}f_{L(R)\sigma}\rangle
+\lambda_{L(R)} \tilde{b}_{L(R)}^2=0
\end{eqnarray}
In order to solve (2) we need to calculate the 
non-equilibrium distribution functions:
$G_{\alpha\sigma, k_{\alpha^{'}}\sigma}^<(t-t')\equiv i\langle 
c^\dagger_{k_{\alpha^{'}}\sigma}(t')
f_{\alpha\sigma}(t)\rangle$ and
$G_{\alpha\sigma, \alpha^{'}\sigma}^<(t-t')\equiv i\langle f^\dagger_{\alpha^{'}\sigma}(t')
f_{\alpha\sigma}(t)\rangle$.
They can be derived by applying the analytic continuation rules of Ref. \cite{Lan} to 
the equation of motion of the time-ordered Green's function
along a complex contour (Keldysh, Kadanoff-Baym or a more general choice).
This allows us to relate $G_{\alpha\sigma,k_{\alpha^{'}}\sigma}^<(t-t')$ with 
$G_{\alpha\sigma, \alpha^{'}\sigma}^<(t-t')$ and 
$G_{\alpha\sigma, \alpha^{'}\sigma}^r(t-t')\equiv -i\theta(t-t')\langle \{f_{\alpha\sigma}(t),
f^\dagger_{\alpha^{'}\sigma}(t')\}\rangle$
and close the set of equations (2) in Fourier space:
\begin{eqnarray}
&&\frac{\tilde{\Gamma}_{L(R)}}{\Gamma}-i
\int\frac{d\epsilon}{2\pi}G_{L,L(R,R)}^<(\epsilon)=\frac{1}{N}\nonumber\\
&&\frac{\tilde{\Gamma}_{L(R)}}{\Gamma}(\tilde{\epsilon}_{L(R)}-
\epsilon_{0})
=i\int\frac{d\epsilon}{2\pi}G_{L,L(R,R)}^<(\epsilon)
(\epsilon-\tilde{\epsilon}_{L(R)}),
\end{eqnarray}
with $\tilde{\epsilon}_{\alpha}=\epsilon_{0}+\lambda_\alpha$ and 
$\tilde{\Gamma}_\alpha=\tilde{b}_\alpha^2\Gamma$. 
For $t_C=0$ , $\tilde{\epsilon}_{\alpha}$ and 
$\tilde{\Gamma}_\alpha$ give, respectively, the position and the width 
of the Kondo peaks in the dot $\alpha$
(at equilibrium, and in the Kondo regime, $\sqrt{\tilde{\epsilon}^2_{\alpha}+\tilde{\Gamma}^2_\alpha}\equiv T^{0}_K=D 
e^{-\pi|\epsilon_0|/\Gamma}$).\cite{Hew} 
The distribution functions in the QD's are: 
$G_{L,L(R,R)}^<(\epsilon)=\frac{2i(\tilde{\Gamma}_{L(R)} f_{L(R)}(\epsilon)
[(\epsilon-\tilde{\epsilon}_{R(L)})^2+\tilde{\Gamma}_{R(L)}^2]
+\tilde{\Gamma}_{R(L)} f_{R(L)}(\epsilon)\tilde{t}_C^2)}
{[(\epsilon-\tilde{\epsilon}_{L}+i\tilde{\Gamma}_L)
(\epsilon-\tilde{\epsilon}_{R}+i\tilde{\Gamma}_R)-\tilde{t}_C^2]
[(\epsilon-\tilde{\epsilon}_{L}-i\tilde{\Gamma}_L)
(\epsilon-\tilde{\epsilon}_{R}-i\tilde{\Gamma}_R)-\tilde{t}_C^2]}$.
$f_{L(R)}(\epsilon)$ is the Fermi-Dirac function in the left (right) lead.
Note that the presence of $\tilde{t}_C^2$ in the denominators indicates that the interdot 
tunneling enters nonperturbatively in the calculations and, then, coherent effects
between dots are fully included. 
Due to the interdot tunneling, the Kondo singularities of each dot 
at $\tilde{\epsilon}_L$ and $\tilde{\epsilon}_R$
combine into coherent superpositions at
$\epsilon_{\pm}=\frac{1}{2}\{(\tilde{\epsilon}_L+\tilde{\epsilon}_R)
\pm\sqrt{(\tilde{\epsilon}_L-\tilde{\epsilon}_R)^2+4\tilde{t}_C^2}\}$.
Of course, at equilibrium 
$\tilde{b}_L=\tilde{b}_R=\tilde{b}$, $\lambda_L=\lambda_R=\lambda$, we recover the 
results of Refs.\cite{Aon,Geor}.
Note that the formation of coherent superpositions of the Kondo singularity is not
trivially related with
its single-particle counterpart (formation of bonding and antibonding states at 
$\epsilon_0\pm t_C$).
Let's focus for simplicity in the equilibrium case 
($\tilde{\epsilon}_L=\tilde{\epsilon}_R$), the splitting is given by 
$\delta\equiv\epsilon_{+}-\epsilon_{-}=2\tilde{t}_C$ which is a 
{\it many-body} parameter
(given by the strong renormalization of the interdot tunneling due to the Kondo effect). 
$\delta$ depends {\it non-linearly} on the single-particle splitting $\delta_0=2t_C$ 
(see Inset of Fig. 3a). In the Kondo limit,
$\{[(\tilde{\epsilon}+\tilde{t}_C)^2+\tilde{\Gamma}^2]
[(\tilde{\epsilon}-\tilde{t}_C)^2+\tilde{\Gamma}^2]\}^{1/4}=T^{0}_K 
e^{\frac{\pi t_C}{\Gamma} (\frac{\tilde{\Gamma}}{\Gamma}-\frac{1}{2})}$.
From the solution of Eq. 3 we obtain the current
$I=\frac{2e}{\hbar} Re\{\sum _{ k_{L}, \sigma}\tilde{V}_L G_{L\sigma,k_L\sigma}^<(t,t)\}$ and
DOS in each QD:
$\rho_{L(R)}(\epsilon)=-\frac{1}{\pi}Im\{\frac{\tilde{b}^2_{L(R)}(\epsilon-\tilde{\epsilon}_{R(L)}
+i\tilde{\Gamma}_{R(L)})}
{[(\epsilon-\tilde{\epsilon}_{L}+i\tilde{\Gamma}_L)
(\epsilon-\tilde{\epsilon}_{R}+i\tilde{\Gamma}_R)-\tilde{t}_C^2]}\}$.\\
{\it Results}:
We solve numerically (for T=0) the set of non-linear equations (3) for 
different voltages $\mu_L=V/2$ and $\mu_R=-V/2$,
$\epsilon_0=-3.5, D=60$ (Kondo regime with $T_K^0\simeq 10^{-3}$)
and different values for the rest of parameters 
(all energies in units of $\Gamma$).
Depending on the ratio $\tau_c=t_C/\Gamma$ we find two different physical scenarios for $\tau_c<1$
and $\tau_c\geq 1$.
\begin{figure}[h]
\centerline{\epsfxsize=0.45\textwidth
\epsfbox{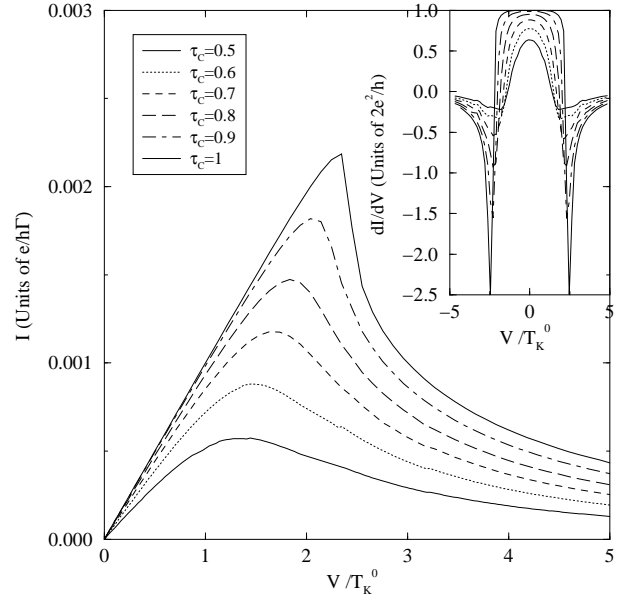}
}
\caption{I-V curves for different values of $\tau_c\leq 1$ and $\epsilon_0=-3.5$.
Caption: dI/dV curves for the same parameters.}
\label{fig:schematic1}
\end{figure}
In Fig. 1 we plot the I-V curves (for clarity, we show only the $V\ge 0$ region) for 
$\tau_c\leq 1$. The two main features
of these curves are: i) An increase of the linear conductance ${\cal G}=dI/dV|_{V=0}$ 
as $\tau_c$ increases;
ii) a saturation, followed by a drop, of the current for large voltages. 
This drop sharpens as $\tau_c\rightarrow 1$.
These features are more pronounced
in a plot of the dI/dV (inset of Fig. 1). 
As $\tau_c$ increases, the zero-bias 
anomaly (originating from the Kondo resonance in the DOS of the dots) becomes broader and 
broader until it saturates into a flat region of value 
$2e^2/h$ (unitary limit)
for $\tau_c=1$. 
The reduction of the current at larger V is reflected as NDC regions in 
the dI/dV curves. For $\tau_c=1$ this NDC becomes singular.
For $\tau_c>1$, and contrary to the previous case, ${\cal G}$ 
{\it decreases} for increasing values of $\tau_c$ (Fig. 2a).
This reduction of ${\cal G}$ can be atributed to the formation of 
the coherent superposition of the Kondo states. 
This can be clearly seen as a splitting $\Delta=2\delta$ in the dI/dV curves (Fig. 2c):
Increasing $\tau_c$, the zero-bias conductance decreases
whereas two maxima at $\pm V_{peak}$ show up (the arrow shows the splitting $\Delta=2V_{peak}$ 
for the maximum
value of $\tau_c$ in the figure).
Fig. 2c demonstrates that the dI/dV curves of a DQD in the Kondo regime directly 
measure the
{\it coherent} combination between the two many-body states in the QD's. 
For larger voltages, the sharp drop of the current (Fig. 2a) reflects as
strong NDC singularities in 
the dI/dV curves 
(Fig. 2b). 
\begin{figure}[h]
\centerline{\epsfxsize=0.45\textwidth
\epsfbox{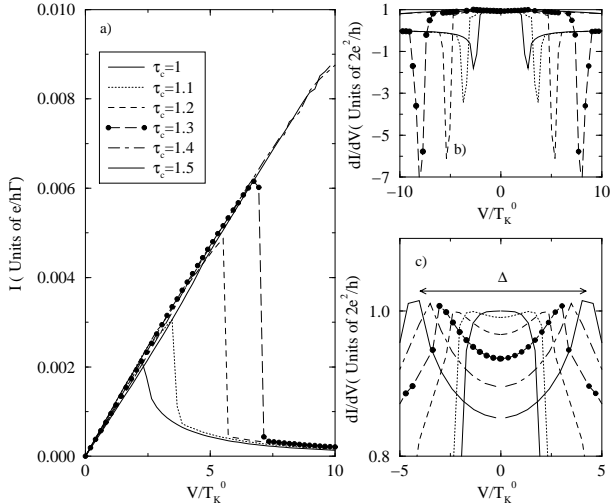}
}
\caption{a) I-V curves for different values of $\tau_c\geq 1$ and
$\epsilon_0=-3.5$.
b) dI/dV curves for the same parameters. c) Blow up of Fig. 2b. The arrow shows the splitting
$\Delta=2\delta$
for $\tau_c=1.5$.}
\label{fig:schematic1}
\end{figure}
The position of these singularities moves towards 
higher $|V|$ as $\tau_c$ increases. 
In order to explain the results of Figs. 1 and 2, we plot
in Fig. 3a $\epsilon_{\pm}$ as a function of $V\ge 0$ for different 
values of $\tau_c$. 
For $\tau_c=0$ (thick solid line) , this corresponds to a plot of 
$\tilde{\epsilon}_{L}$ and $\tilde{\epsilon}_{R}$
(i.e the positions of the Kondo resonances for the decoupled QD's) 
as a function of $V$.
We obtain, as expected, that each Kondo resonance is pinned at the chemical potential
of its own lead, $\tilde{\epsilon}_{L}=\mu_L=V/2$ and $\tilde{\epsilon}_{R}=\mu_R=-V/2$. 
As the interdot coupling 
is turned on, the voltage dependence becomes strongly non-linear.
For low V, the curves for $\tau_c\neq 0$ do not coincide with the curves
for $\tau_c=0$ (i.e, $\mu_{L/R}$). This situation, however, changes as we increase 
$V$; the level positions $\epsilon_\pm$ converge towards
the chemical potentials $\mu_{L/R}$ in a non-trivial way.
\begin{figure}[h]
\centerline{\epsfxsize=0.45\textwidth
\epsfbox{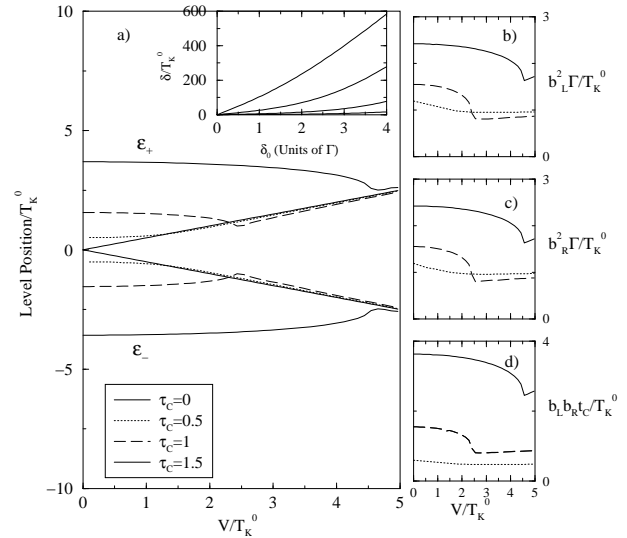}
}
\caption{a)
$\epsilon_{\pm}$ vs. V
for different values of $\tau_c$ ($\epsilon_0=-3.5$).
Inset: Many body splitting ($\delta$)
as a function of the single particle splitting ($\delta_0$)
for $V=0$ and $\epsilon_0=-2.0,-2.5,-3.0,-3.5$
(from top to bottom).
b) $\tilde{\Gamma}_L$ vs. V.
c) $\tilde{\Gamma}_R$ vs. V.
d) $\tilde{t}_C$ vs. V.}
\label{fig:schematic1}
\end{figure}
The voltage $V$ for which 
$\epsilon^+-\epsilon^-$ coincides with the chemical potential difference $V$
 gives 
the position of the peak in the positive side of the dI/dV (Fig.2c).
This voltage is the solution of the  equation
$\delta(V_{\rm peak})=V_{\rm peak}$ where  
$\delta(V)\equiv \sqrt{(\tilde{\epsilon}_L-\tilde{\epsilon}_R)^2
+4\tilde{t}_C^2}$,
with $\tilde{\epsilon}_{L/R}$ given by Eq. (3). Note the implicit
(and non-trivial) voltage dependence of 
$\delta(V)$.
$\tilde{\Gamma}_L$, $\tilde{\Gamma}_R$ and $\tilde{t}_C$
follow a similar behavior as a function of $V$ (Figs. 3b, 3c and 3d).
For $V\geq V_{\rm peak}$, we find numerically that
$\delta(V)\approx V$, a relationship that becomes asymptotically exact
as $V\rightarrow\infty$.
The equation $\delta(V)=V$,
has stable solutions $\tilde{t}_C\neq 0$ for 
$[\frac{(\tilde{\epsilon}_L-\tilde{\epsilon}_R)}{V}]^2<1$,
while for $[\frac{(\tilde{\epsilon}_L-\tilde{\epsilon}_R)}{V}]^2>1$,
the only stable solution  is $\tilde{t}_C=0$, corresponding to
current $I=0$.  We denote the crossover voltage where
$[\frac{(\tilde{\epsilon}_L-\tilde{\epsilon}_R)}{V}]^2=1$ by $V^*$.
For finite voltages $V>V_{\rm peak}$, on the other hand, the relation
$\delta(V)=V$ is only approximate, so that at the crossover
$\approx V^*$, the quantity $\tilde{t}_C$ and hence $I$ drop to
a much smaller, but still finite, value instead.  Nevertheless the crossover
at $V\approx V^*$ still indicates the beginning of the NDC region.

To illustrate this, we plot in Fig. 4 the left and right QD's DOS for $\tau_c=1$.
At equilibrium (V=0), the Kondo singularity at $\epsilon=0$ splits into
the $\epsilon_{\pm}$ combinations.
For $V/T_K^0=2$ the coherence is still preserved but
the physical picture utterly changes for higher voltages
($V/T_K^0=4$ and $V/T_K^0=6$).
In this case, the previous configuration is no longer stable,
the coherence between dots is lost ($\tilde{t}_C\rightarrow 0$), the dots are almost decoupled
and the Kondo resonances in each dot
are pinned again at their own chemical potential:
the weight of the left (right) DOS at $\epsilon=\mu_{R(L)}$ is almost zero (even though $\tau_c=1$).\\
This instability resembles that of the SB at
$T\neq 0$ in the single impurity Anderson hamiltonian. \cite{Gauss,Piers} 
In the MFA the SB behaves as the order parameter
associated with the conservation of $Q$.
When $\tilde{b}\neq 0$ the gauge
symmetry $b\rightarrow b e^{i\theta}$, $f\rightarrow f e^{i\theta}$
associated with charge conservation is broken and the MFA has two phases
$\tilde{b}\neq 0$ and $\tilde{b}=0$
separated by a second order phase transition.
It is important to point out that the fluctuations
do not destroy completely this $\tilde{b}\rightarrow 0$ behavior
(the SB fluctuations develop power law behavior
replacing the transition by a smooth crossover).
We speculate that in our problem this zero-temperature transition at finite V may be 
also robust against fluctuations 
but $1/N$ corrections are needed to substantiate this argument. Work in this direction is in progress.
\begin{figure}[h]
\centerline{\epsfxsize=0.45\textwidth
\epsfbox{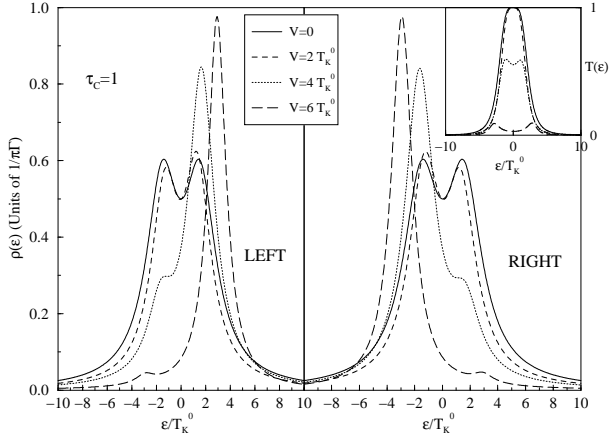}
}
\caption{DOS for the left and right dot for $\tau_c=1$ and different voltages 
($\epsilon_0=-3.5$). Caption: Transmission probability of the DQD for the same parameters.}     
\label{fig:schematic1}
\end{figure}

In closing, we have demonstrated that the non-linear 
transport properties $(dI/dV)$ of a DQD in the Kondo regime
directly measures the transition (as $t_C$ increases)
from two isolated Kondo impurities
to a coherent bonding and antibonding
superposition of the many-body Kondo states of
each dot. While for $t_C<\Gamma$ the conductance maximum is
at $V=0$, for $t_C>\Gamma$ the transport is optimized
for a finite $V$ matching the splitting between these two bonding
and antibonding states.
For large voltages (and $t_C\geq \Gamma$)
there is a critical voltage above which
the coherent superposition is unstable and the physical behavior of the system
again resembles that of two decoupled QD's. This leads 
a strong reduction of the current and
singular regions of negative differential conductance.
Concerning the observability of these effects: In our MFA
the maximum value of $\delta$ ranges from 
$\delta\simeq 20 T_K^0-500 T_K^0$ (inset of Fig. 3a)
giving,
for the experiment of Ref. \cite{Gold} ($\Gamma\sim 150\mu eV$),
$\delta\sim 3\mu eV-75\mu eV$ (30mK-750mK) which is within the resolution limits of present day techniques.\\
This work was supported by the NSF
grant DMR 97-08499, DOE grant DE-FG02-99ER45970 and by the MEC of Spain grant PF 98-07497938.


\begin{references}
\bibitem{wide}
$\Gamma(\epsilon)=\pi\sum_{k_{\alpha}}|V_0|^2\delta(\epsilon-\epsilon_{k_\alpha})$.
We take $\Gamma\equiv\Gamma(\epsilon_F)$ for $-D\le \epsilon\le D$
($D$ serves as a cutoff in the calculations).
\bibitem{inter}
The neglect of $U_{\rm interdot}$ corresponds to the experimentally
accessible limit of small interdot capacitance as compared with
the capacitances of each QD to the gates, and implies
a vanishing interdot antiferromagnetic coupling from this source.
The antiferromagnetic coupling due to superexchange also vanishes
in the model, because $U_{\rm intradot}\rightarrow\infty$.
\end{references}
\end{document}